\documentclass[12pt,letterpaper]{JHEP3}

\usepackage{amsmath,amssymb}
\usepackage{graphicx}

\newcommand{\be}{\begin{equation}}
\newcommand{\ee}{\end{equation}}
\newcommand{\ba}{\begin{aligned}}
\newcommand{\ea}{\end{aligned}}
\newcommand{\ben}{\begin{displaymath}}
\newcommand{\een}{\end{displaymath}}
\newcommand{\bea}{\begin{eqnarray}}
\newcommand{\eea}{\end{eqnarray}}
\newcommand{\bean}{\begin{eqnarray*}}
\newcommand{\eean}{\end{eqnarray*}}

\newcommand{\p}{\partial}

\def\l {\lambda}

\def\a {\alpha}

\def\b {\beta}

\def\g {\gamma}

\def\s {\sigma}

\title{\Large Giant Magnons in Symmetric Spaces:\\
Explicit $N$-soliton solutions for $CP^n$, $SU(n)$ and $S^n$}

\author{Chrysostomos Kalousios~$^{a}$ and Georgios Papathanasiou~$^{b}$

\\$^{a}$Institut f\"ur Physik der Humboldt-Universit\"at zu Berlin,
Newtonstra{\ss}e 15, D-12489 Berlin, Germany \\
$^{b}$Brown University, Providence, Rhode Island 02912, USA\\
E-mail: \email{ckalousi@physik.hu-berlin.de}, \email{Georgios\_Papathanasiou@brown.edu}}

\abstract{
Giant magnons are one of the main manifestations of integrability on the string theory side of the AdS/CFT correspondence. Motivated by the recent advances in their study, especially in the context of the string theory dual of ABJM theory, we present and prove explicit $N$-soliton solutions for the relevant $CP^n$, $SU(n)$ and $S^n$ sigma models. The proof is based on solving the dressing method recursion with the help of determinant operations, and our solutions hold for any choice of vacuum and soliton parameters. We further specialize our results for the choices that lead to giant magnons, and as an application, we calculate the classical time delay due to the scattering of an arbitrary number of $CP^2$ elementary dyonic magnons. The determinant expressions for our $N$-soliton solutions could possibly be used for the derivation of an effective particle description of magnon scattering.}


\preprint{\small{HU-EP-10/13}\\
\small{Brown-HET-1595}}

\begin{document}

\section{Introduction}

Integrable structures have been quintessential in the spectroscopic analysis of gauge-string dualities. First discovered in the prime example of the $AdS_5/CFT_4$ correspondence \cite{Maldacena:1997re} between $\mathcal{N}=4$ super-Yang Mills theory and Type IIB strings on $AdS_5\times S^5$ (see the reviews \cite{Beisert:2004ry, Zarembo:2004hp, Plefka:2005bk, Rej:2009je} and references therein), these structures also appear in the more recent $AdS_4/CFT_3$ case, between the ${\cal{N}} = 6$ superconformal Chern-Simons theory of \cite{Aharony:2008ug} and Type IIA strings on $AdS_4\times CP^3$ \cite{Minahan:2008hf, Bak:2008cp, Gaiotto:2008cg, Gromov:2008qe, Ahn:2008aa}. In the gauge theory side, anomalous dimensions of local operators can be calculated using the Bethe ansatz of an integrable spin chain.

In the string theory side, the language of spin chains is translated into the framework of integrable sigma models \cite{Pohlmeyer:1975nb, Zakharov:1973pp, Harnad:1983we} which arise in various subsectors, and especially their solitonic solutions. For $AdS_5/CFT_4$, great progress towards establishing a precise connection was made when a single spin chain excitation in a certain asymptotic limit, was mapped to its strong coupling equivalent, a classical string solution termed giant magnon \cite{Hofman:2006xt}. It is a solitonic string in the $R\times S^2$ subspace of $AdS_5\times S^5$, with its endpoints moving on an equator of $S^2$ with the speed of light. A dyonic generalization of the original solution, which moves on $R\times S^3$ with an additional conserved charge, was found in \cite{Dorey:2006dq, Chen:2006gea}, and a large family of giant magnons has been further constructed by nonlinear superposition of the aforementioned solutions \cite{Spradlin:2006wk, Kalousios:2006xy, Kalousios:2008gz}. At the classical level, giant magnons have been very successful in verifying the all-loop proposals for the dispersion relation and the S-matrix of the dual gauge theory \cite{Beisert:2005tm, bes}, with similar success also in the next order in the semiclassical expansion \cite{Papathanasiou:2007gd, Chen:2007vs}.

It is due to their contribution in confirming interpolating quantities at strong 't Hooft coupling, that similar giant magnon solutions on $R\times CP^3$ have been the subject of intense study in the $AdS_4/CFT_3$ context. As was pointed out in \cite{Abbott:2008qd}, the initial attempts \cite{Gaiotto:2008cg, Grignani:2008is} were different embeddings of known solutions to various spherical subspaces of $CP^3$, whereas new breather-like solutions were found in \cite{Hollowood:2009tw, Kalousios:2009mp, Suzuki:2009sc}. More importantly, a new dyonic solution was discovered, first for a special kinematic configuration in \cite{Abbott:2009um}, and then for the general case in \cite{Hollowood:2009sc}, which is elementary\footnote{This is not to be confused with the notion of elementary excitations in the spin chain of the dual CFT. Due to its dyonic nature, this giant magnon is dual to BPS bound states of many spin chain magnons, as was the case for \cite{Chen:2006gea}.} in the sense that all previously known solutions are either composites or limits of it. By calculating the classical phase shift from the scattering of two such dyonic magnons, Hatsuda and Tanaka \cite{Hatsuda:2009pc} performed a nontrivial strong coupling test of the all-loop proposal for the $AdS_4/CFT_3$ S-matrix\footnote{Subtleties of the $AdS_4/CFT_3$ integrability, when contrasted with $AdS_5/CFT_4$, have been pointed out in \cite{McLoughlin:2008ms, Alday:2008ut, Krishnan:2008zs}.} \cite{Ahn:2008aa}.

Inspired by these developments, in this paper we look at the problem of constructing explicit $N$-soliton solutions for sigma models with target space a manifold $M$, which when supplemented with Virasoro constraints, are equivalent, in the conformal gauge, to string theory on a $R\times M$ subsector of the backgrounds that are relevant for the $AdS_d/CFT_{d-1}$ correspondence. In particular, we will first consider the $M=CP^n$ case, which as we described arose more recently in the $d=4$ instance of the correspondence, and then generalize our results for $M=SU(n)$ and $S^n$, related to the $d=5$ instance\footnote{As we explain in appendix \ref{appSU2}, strings on $S^3$ can be embedded in the $SU(2)$ sigma model.}.

The main tool for our calculation will be the dressing method of Zakharov and Mikhailov \cite{Zakharov:1973pp,Zakharov:1980ty}, first developed for solving the sigma model with target space $SL(n,C)$ or any of its subgroups $G$, such as $SU(n)$ or $SO(n)$, and further refined in \cite{Harnad:1983we} so as to hold for all submanifolds that are symmetric spaces $G/H$, such as the $CP^n=SU(n+1)/U(n)$ and $S^n=SO(n+1)/SO(n)$ cases of interest\footnote{In fact, all subgroups can be given a symmetric space structure, so that the analysis of the latter reference encompasses them.}. It is a systematic recursion whereby starting from a known, usually simple, solution which is called the vacuum, a new solution is produced at each step, from the previously determined one. Our derivation is based on solving the dressing method recursion, first by expressing each step of the recursion in determinant form, and then proving the equality of different determinants by means of elementary row and column operations.

The $N$-soliton solutions we present are very general, in the sense that they hold for any choice of vacuum and individual soliton parameters, which as we will see consist of a complex number whose argument is proportional to the soliton momentum, and a polarization vector. Hence, although in our analysis the paradigm has been giant magnons, for which we specialize our results for the appropriate choices of vacua, our formulas can be used with various other choices in order to produce different solitonic solutions. For example, it is known that the dressing method can produce spiky strings \cite{Ishizeki:2007kh}, so the vacuum choice of the latter reference will provide us with $N$-spike solutions.

As an application, we move on to calculate the classical time delay due to the scattering of an arbitrary number of $CP^2$ dyonic magnons. In this manner, we verify that the magnon polarizations which were chosen to correspond to spin chain excitations in \cite{Hatsuda:2009pc}, indeed lead to factorized scattering as expected from the gauge theory. Apart from the scattering solutions that arise for generic values of the soliton parameters, it is worth mentioning that according to the analysis of \cite{Hollowood:2009tw}, particular values may also lead to a combination of elementary solitons and breathers. This is analogous to the $N$-soliton solutions of the sine-Gordon equation, whereby two solitons are replaced by a breather when their parameters are taken to be conjugate to each other \cite{Babelon:1993bx}.

As a final remark  about the utility of our results, we should note that knowledge of the $N$-soliton solution serves as a useful starting point in deriving an effective particle Hamiltonian description of the sine-Gordon theory, so it is likely that similar techniques may prove useful in the case of generalized sine-Gordon theories and sigma models.  Positions and momenta in the so called Ruijsenaars-Schneider Hamiltonian \cite{Ruijsenaars:1986vq} are related to the positions and rapidities of the sine-Gordon solitons \cite{Babelon:1993bx}, and the phase shift for soliton scattering can be calculated from the quantum mechanical model.  Interest of an effective particle description of giant magnons emerged through the work of \cite{Dorey:2007xn} (see also \cite{Aniceto:2008pc,Dorey:2010ag}).

Our paper is organized as follows.  In chapter 2 we start with a review of the dressing method for the $SL(n,C)$ principal chiral model and its symmetric space reductions, and then focus on the particular case of $CP^n$. As a warmup we reproduce the 1- and 2-soliton solutions, and most importantly build up our notations for handling the $N$-th step of the dressing method recursion. In chapter 3, we present and prove our $N$-soliton formulas, initially for the $CP^n$ and consecutively for the $SU(n)$ and $S^n$ sigma models. Chapter 4 deals with the application of our results in the calculation of the classical time delay due to the scattering of an arbitrary number of $CP^2$ dyonic magnons. Chapter 5 includes our conclusions and possible directions of further inquiry. Finally, in appendix A we remind the reader of the definition of the symmetric spaces, and in appendix B we give examples of giant magnon solutions for specific symmetric space dimensions, soliton polarizations and/or vacua, and work out the final expressions for the $N$-magnon formulas in these cases of interest.

\section{Sigma Models and Dressing}
We start with an $SL(n,C)$ field $g$  with an action proportional to
\be
\int d^2\s \text{Tr}[(\partial_a g g^{-1})(\partial^a g g^{-1})],
\ee
with the equations of motion having the form of a conservation law,
\be\label{s-model_EOM}
\partial_a J^a=0,\quad J^a=\partial^a g g^{-1},
\ee
for the current $J^a$. We will be working in light-cone coordinates,
\be
x_\pm = \frac 1 2 (t \pm x),\quad\p_\pm =\p_t \pm \p_x,
\ee
where $t,x$ are the Minkowski coordinates in the worldsheet space, $\eta^{ab}=\text{diag}(-1,1)$.

The dressing method \cite{Zakharov:1973pp,Zakharov:1980ty} is a systematic method for obtaining new solitonic solutions of the classically integrable equation (\ref{s-model_EOM}), by a nonlinear superposition of known ones. This is done by mapping the nonlinear equation in question, to a system of linear equations for the auxiliary field $\Psi(x;\l)$
\be\label{linearsystem}
\p_\pm \Psi(x;\l)= \frac{J_\pm \Psi(x;\l)} {1 \pm \l},
\ee
which is required to hold for all values of the complex spectral parameter $\l$ that it introduces, with $J_\pm$ independent of it, and $\Psi(x;\l)$ invertible. Indeed these requirements imply that if $\Psi(x;\l)$ is a solution of (\ref{linearsystem}), then $\Psi(x;0)$ is a solution of (\ref{s-model_EOM}).

The advantage of the auxiliary system is that it is invariant under a $\l$-dependent gauge transformation (from now on we drop the $x$-dependence)
\be
\Psi(\l)\rightarrow \Psi'(\l)=\chi(\l)\Psi(\l),
\ee
with the $J_\pm$ also transforming appropriately. The dressing factor $\chi(\l)$ will depend on $\Psi(\l)$, and its precise form is determined by the requirement that the transformed $J'_\pm$ remain independent of $\l$, together with analyticity properties which we will refer to shortly.

The symmetries of the auxiliary system allow us to construct new, nontrivial solutions to (\ref{s-model_EOM}) in the following manner: We start with a known solution\footnote{This is called the vacuum solution, and usually corresponds to constant $J_\pm$.}  $g_0$ of (\ref{s-model_EOM}), and solve (\ref{linearsystem}) with the boundary condition $\Psi_0(0)=g_0$ and $\Psi_0(\infty)=I$. Then, we use the gauge transformation to obtain a new $\Psi_1(\l)=\chi_1(\l)\Psi_0(\l)$, which also leads to a new solution $g_1=\chi_1(0)g_0$ of (\ref{s-model_EOM}). More solutions may be produced by multiplying each obtained $\Psi_i(\l)$ with a new dressing factor, for example $\Psi_2(\l)=\chi_2(\l)\Psi_1(\l)$. Note that we only need to solve differential equations in the first step, and then the method proceeds in a purely algebraic manner.

The virtue of the aforementioned framework is that it not only holds for the $SL(n,C)$ sigma model, but also for a large class of integrable systems obtained by restricting to suitable submanifolds of the target space. Such submanifolds can be either subgroups $G$ of $SL(n,C)$ \cite{Zakharov:1973pp}, or symmetric spaces $G/H$ contained in it \cite{Harnad:1983we}, namely quotient spaces with some additional structure, which is explained in more detail in appendix \ref{appx_symm_spaces}. In all cases the dressing factor $\chi(\l)$ and its inverse are taken to be meromorphic functions of $\l$,
\be\label{chi_meromorphic}
\chi(\l)=1+\sum_{i=1}^K\frac{Q_i}{\l-\l_i}\;,\qquad\chi^{-1}(\l)=1+\sum_{i=1}^K\frac{R_i}{\l-\l_i}.
\ee
Analyticity places further constraints on the residues $Q_i, R_i$ and allows for the determination of their general structure, provided in Theorem 4.2 of \cite{Harnad:1983we}. Additional input is required for the rank of the residues, and for the remaining of our paper, we will study the simplest case where
\be
\text{rank}(Q_i)=\text{rank}(R_i)=1.
\ee
What certainly differs for each symmetric space $G/H$ is how it is embedded in a submanifold $\Sigma$ of $SL(n,C)$, or conversely the constraints the $SL(n,C)$ matrix field $g$ should obey in order to lie in $\Sigma$. They are distinguished in two categories, those that involve a function $\s_+$ and restrict $g \in G$, and those that involve the function $\s_-$ of the symmetric space structure,
\be
\Sigma=\{g\in SL(n,C)|\s_+(g)=g, \s_-(g)=g^{-1}\}.
\ee
All possible symmetric spaces in $SL(n,C)$ are presented in Table 1 of \cite{Harnad:1983we}, together with the precise form of the functions $\s_\pm$ for each case.

The constraints on $g$ can be further extended to $\Psi(\l)$ and $\chi(\l)$, and in particular imply a minimum number of poles in (\ref{chi_meromorphic}), $K=K_{min}$, as well as relations between their positions, also included in the aforementioned Table. Since it can be shown that dressing transformations with more poles simply correspond to a sequence of transformations with fewer poles, in this paper we will only consider $K=K_{min}$. Finally, to fully fix the form of $\chi(\l)$, one needs to derive additional relations between the constituents of the residues $Q_i, R_i$, implied by the constraints. These have been summarized in Table 2 of \cite{Harnad:1983we}.

\subsection{Single soliton}

For the remainder of this chapter we will focus on the sigma model with a $CP^n = SU(n+1) / U(n)$ target space \cite{Sasaki:1984tp}. This is described by a $(n+1)$-component vector $Z$, $Z\sim \mu Z$ for any nonzero complex number $\mu$, which may be embedded in $SU(n+1)$ by defining the matrix field
\be\label{P0}
g=\theta \left( 1-2P \right), \quad P=\frac{Z Z^\dagger}{|Z|^2}, \quad \theta={\rm diag}(-1,1,\ldots,1),
\ee
where evidently $P$ is a hermitian projector operator, i.e. it satisfies $P^2=P=P^\dagger$, of rank 1.

As we have mentioned in the introduction, our motivation for studying this model is its equivalence to string theory on $R\times CP^n$ in the conformal gauge, and in particular the fact that for $n=3$ it emerges as a subsector of the dual to ABJM theory, in the context of $AdS_4/CFT_3$ duality \cite{Aharony:2008ug}. The corresponding Virasoro constraints are
\be\label{Virasoro44}
{\rm Tr} [J_\pm^2] = {\rm Tr} [(\p_\pm g g^{-1})^2] = -2\kappa^2,
\ee
coming from the gauge choice $T=\kappa t$, where $T$ is the $AdS$ global time coordinate. From Table 1 of \cite{Harnad:1983we}, or \cite{Sasaki:1984tp}, we infer that the constraints on $g$ are
\be\label{gconstraints}
g g^\dagger=g\theta g\theta =1,
\ee
which further imply the subgroup and coset constraint for $\Psi(\l)$ respectively,
\be
\qquad [\Psi(\bar{\l})]^\dagger \Psi(\l) =1,\quad \Psi(\l) = \Psi(0) \theta \Psi (1/ \l) \theta.
\ee
Starting from a vacuum solution $g_0$, we may obtain a 1-soliton solution\footnote{In what follows we will be using the terms `soliton' and `magnon' interchangeably, keeping in mind that the latter is a special case of the former, for a particular choice of vacuum.} $g_1=\Psi_1(0)$ from $\Psi_1(\l)=\chi_1(\l)\Psi_0(\l)$, where the dressing matrix $\chi_1(\l)$ is given by
\be\label{chi_1}
\chi_1(\lambda)=1+\frac{Q_1}{\l-\xi_1}+\frac{Q_2}{\l-1/\xi_1},
\ee
and the matrices $Q_1, Q_2$ by
\be\ba\label{Q12}
Q_1&=\frac 1 \Delta \left( -\frac{\xi_1 \bar{\xi}_1 \beta_{11}}{\xi_1-\bar{\xi}_1} \theta h_1 h_1^\dagger \theta +\frac{\xi_1 \gamma_{11}}{\xi_1 \bar{\xi}_1-1}g_0 h_1 h_1^\dagger \theta \right),\\
Q_2&=\frac 1 \Delta \left( \frac{\beta_{11}}{\xi_1-\bar{\xi}_1} g_0 h_1 h_1^\dagger g_0^\dagger -\frac{\bar{\xi}_1 \gamma_{11}}{\xi_1 \bar{\xi}_1-1} \theta h_1 h_1^\dagger g_0^\dagger \right).
\ea\ee
In the above we have defined the convenient quantity
\be\label{cph}
h_i=\theta \Psi_0(\bar{\xi}_i) e_i,
\ee
where for the moment $i=1$, but as we will see in the next sections, in general $i$ can vary up to the total number $N$ of solitons  of our solution. The complex $(n+1)$-dimensional vector $e_i$ is called the polarization vector of the $i$-th soliton and can be arbitrary, whereas the complex numbers $\beta_{ij},\gamma_{ij}$ are defined to be
\be\label{cpbeta}
\beta_{ij}=h_i^\dagger h_j, \quad \gamma_{ij}=h_i^\dagger \theta g_0 h_j,
\ee
and the real number $\Delta$ by
\be\label{Delta}
\Delta=-\frac{\xi_1 \bar{\xi}_1 \beta_{11}^2}{(\xi_1-\bar{\xi}_1)^2}+\frac{\xi_1 \bar{\xi}_1 \gamma_{11}^2}{(\xi_1 \bar{\xi}_1-1)^2}.
\ee

Although the dressing method generically holds for the matrix fields $g$, it is also possible to formulate it directly in terms of the $CP^n$ target space coordinate vectors $Z$, for which it takes a simpler form \cite{Sasaki:1984tp}. For the single magnon example we are examining, we do this by finding the eigenvector $Z_1$ of the rank one projection operator $P_1=\frac 1 2 (1-\theta g_1)$, after we replace $g_1=\chi_1(0)g_0$. From \eqref{P0}, the useful identities
\be
h_i^\dagger P_0 h_j=\frac{(\beta-\gamma)_{ij}}{2}, \qquad P_0 h_ih_j^\dagger P_0=\frac{(\beta-\gamma)_{ji}}{2}P_0
\ee
we can derive from it, and \eqref{chi_1}-\eqref{Q12}, we can express $P_1$ as
\be\label{P1}
P_1=\frac 1 \Delta \left(\alpha_{11} \bar \alpha_{11}P_0+\frac{(\beta-\gamma)_{11}}{2}h_1 h_1^\dagger+\alpha_{11} P_0 h_1 h_1^\dagger+\bar{\alpha}_{11} h_1 h_1^\dagger P_0 \right),
\ee
where
\be\label{cpalpha}
\alpha_{ij}=-\frac{\xi_i \beta_{ij}}{\xi_i-\bar{\xi}_j}-\frac{\gamma_{ij}}{\xi_i \bar{\xi}_j-1}\;.
\ee
The latter form of $P_1$ is useful because if we define
\be\label{Z0dressing}
Z_1=(\alpha_{11}+h_1 h_1^\dagger) Z_0,
\ee
it immediately follows that
\be
Z_1 Z_1^\dagger=\Delta |Z_0|^2 P_1,
\ee
whereas the norm of $Z_1$ is given by
\be
|Z_1|^2=\Delta |Z_0|^2.
\ee
Hence
\be\label{P_1}
P_1=\frac{Z_1 Z_1^\dagger}{|Z_1|^2}\Rightarrow P_1 Z_1=Z_1,
\ee
so that $Z_1$ defined in \eqref{Z0dressing} indeed corresponds to the coordinate vector of the single magnon solution.

As an added bonus from this procedure, we obtain an alternative expression for $\Delta$,
\be\label{Delta_alternative}
\Delta=\frac{|Z_1|^2}{|Z_0|^2}=\a_{11}\bar{\a}_{11}+(\a_{11}+\bar{\a}_{11}+\b_{11})\frac{\b_{11}-\gamma_{11}}{2},
\ee
which will be of use in the next section.

\subsection{Two solitons}

Before proceeding to the analysis of the $N$-soliton solution, it is important to determine the relation between quantities appearing in consecutive dressing factors, for which the study of the 1- to 2-soliton transition can serve as a very instructive example.

In order to obtain $\Psi_2$, we would have to multiply $\Psi_1$ with a dressing factor $\chi_2$ which now has poles at new positions $\xi_2, 1/\xi_2$ rather than $\xi_1, 1/\xi_1$, and whose residues will necessarily depend on $\Psi_1$ rather than $\Psi_0$. From these considerations, it is easy to see that the generalization of \eqref{Z0dressing} for the dressing of $Z_1$ will be given by
\be
Z_2=(\alpha_{22}'+h_2'h_2'^\dagger)Z_1,
\ee
where we define
\be
h_i'=\theta \Psi_1(\bar{\xi}_i) e_i = \theta \chi_1(\bar{\xi}_i) \theta h_i
\ee
and
\be\label{primeabc}
\alpha_{ij}'=-\frac{\xi_i \beta_{ij}'}{\xi_i-\bar{\xi}_j}-\frac{\gamma_{ij}'}{\xi_i \bar{\xi}_j-1}, \quad \beta_{ij}'=h_i'^\dagger h_j', \quad \gamma_{ij}'=h_i'^\dagger \theta g_1 h_j'.
\ee
As we stated in the beginning of this section, the question now is how to express the primed quantities in terms of the unprimed ones. Starting with $h_i'$, we can do this by replacing $\chi_1$ from \eqref{chi_1}-\eqref{Q12} and also using $\theta g_0=g_0^\dagger \theta=1-2P_0$. In this fashion, we obtain
\be\ba\label{hiprime}
h_i'&=h_i+c_{1,i}h_1+\tilde c_{2,i}P_0 h_1\\
&=h_i+c_{1,i}h_1+ c_{2,i}Z_0\;,&&c_{2,i}=\tilde c_{2,i} \frac{Z_0^\dagger h_1}{|Z_0|^2},
\ea\ee
where in order to go from the first to the second line, we have just used the definition \eqref{P0} for $P_0$. The coefficients of the vectors are given by
\be\ba
c_{1,i}&=\frac 1 \Delta \left[\frac{\xi_1\beta_{1i}}{\bar{\xi}_i-\xi_1} \left( -\frac{\bar{\xi}_1 \beta_{11}}{\xi_1-\bar{\xi}_1}  +\frac{\gamma_{11}}{\xi_1 \bar{\xi}_1-1}  \right) +\frac{\xi_1 \gamma_{1i}}{\xi_1 \bar{\xi}_i-1}  \left( \frac{\beta_{11}}{\xi_1-\bar{\xi}_1} -\frac{\bar{\xi}_1 \gamma_{11}}{\xi_1 \bar{\xi}_1-1} \right) \right]\\
\tilde c_{2,i}&=-\frac {2} {\Delta}\left[\frac{\xi_1\beta_{1i}}{\bar{\xi}_i-\xi_1} \frac{\gamma_{11}}{\xi_1 \bar{\xi}_1-1} +\frac{\xi_1\gamma_{1i}}{\xi_1 \bar{\xi}_i-1}  \frac{\beta_{11}}{\xi_1-\bar{\xi}_1} \right],
\ea\ee
which with the help of the definition \eqref{cpalpha} can be rewritten as
\be\ba\label{c's}
c_{1,i}&=\frac 1 \Delta \Big[-\a_{1i}\bar{\a}_{11}-(\a_{1i}+\bar{\a}_{i1}+\b_{1i})\frac{\b_{11}-\gamma_{11}}{2}\Big]\\
\tilde c_{2,i}&=\frac 1 \Delta (\a_{1i}\bar{\a}_{11}-\a_{11}\bar{\a}_{i1}+\a_{1i}\b_{11}-\a_{11}\b_{1i}).
\ea\ee
The advantage of the latter relations, when used with the equivalent relation for $\Delta$ \eqref{Delta_alternative}, is that all $\xi_i$ dependence is absorbed in the $\a$'s, which should appear in the final expression. From \eqref{c's} we can also prove the following identity,
\be
c_{1,i}=\frac{\tilde c_{2,i}}{\a_{11}}\frac{\b_{11}-\gamma_{11}}{2}-\frac{\a_{1i}}{\a_{11}}= c_{2,i}\frac{h_1^\dagger Z_0}{\a_{11}}-\frac{\a_{1i}}{\a_{11}},
\ee
which allows us to eliminate $c_{1,i}$ in \eqref{hiprime}, yielding
\be\label{hprime}
h_i'=h_i-\frac{\a_{1i}}{\a_{11}}h_1+\frac{c_{2,i}}{\a_{11}}Z_1.
\ee
Multiplying on the left with $Z^\dagger_1$, replacing $c_{2,j}$ from \eqref{hiprime} and \eqref{c's}, and finally conjugating, produces another important identity between primed and unprimed quantities,
\be\label{h'Z1}
h_i^{' \dagger } Z_1=\alpha_{11}h_i^{\dagger}Z_{0}-\alpha_{i1}h_1^{\dagger}Z_{0}.
\ee
After a lengthy calculation, we may similarly obtain a relation between primed and unprimed $\a$'s,
\be\label{alphaprime}
\alpha'_{ij}=\alpha_{ij}-\frac{\alpha_{i1}\alpha_{1j}}{\alpha_{11}}-\frac{ c_{2,j}}{\alpha_{11} }h_i^{'\dagger} Z_1.
\ee
Let us sketch the main steps for a derivation of (\ref{alphaprime}). The reader interested in its applications can jump straight to the paragraph after (\ref{bg_identity}).  We start by recasting $\b'_{ij}$ with the help of (\ref{hprime}) as
\be\label{b'}
\b'_{ij}=(h_i^\dag-\frac{\bar \a_{1i}}{\bar \a_{11}}h_1^\dag)(h_j-\frac{ \a_{1j}}{ \a_{11}}h_1)+\frac{c_{2,j}}{\a_{11}}h_i^{' \dagger } Z_1+\frac{\bar c_{2,i}}{\bar\a_{11}}Z_0^\dag (\bar\a_{11}+h_1 h_1^\dag)(h_j-\frac{\a_{1j}}{\a_{11}}h_1),
\ee
where we have already substituted $Z_1^\dag$ from (\ref{Z0dressing}) in the last term. We further proceed by reexpressing $h_i^{' \dagger } Z_1$ through (\ref{h'Z1}), and replacing $h_k^\dag h_l=\b_{kl}$ after we expand all parentheses. We also exchange the $c$ coefficient with $\tilde c$ from (\ref{hiprime}), and similarly for ${\bar c}$.

As far as $\gamma'_{ij}$ is concerned, it is advantageous to work with the quantity $(\b'_{ij}-\g'_{ij})/2$ instead, as it can be easily expressed in terms of unprimed quantities through (\ref{P_1}), (\ref{Delta_alternative}) and (\ref{h'Z1}),
\be\label{(b-c)'}
\begin{aligned}
\frac{\b'_{ij}-\g'_{ij}}{2}&=h^{'\dag}_i P_1 h'_j=\frac{1}{|Z_1|^2} h^{'\dag}_i Z_1 Z_1^\dag h'_j\\
&=\frac{1}{\Delta |Z_0|^2}(\alpha_{11}h_i^{\dagger}Z_{0}-\alpha_{i1}h_1^{\dagger}Z_{0})(\bar \alpha_{11}Z_{0}^{\dagger}h_j-\bar \alpha_{j1}Z_{0}^{\dagger}h_1).
\end{aligned}
\ee
We arrive at the final form of both (\ref{b'}) and (\ref{(b-c)'}) after we replace all combinations
\be\label{b-c}
\frac{1}{|Z_0|^2} (h^{\dag}_k Z_0) (Z_0^\dag h_l)=\frac{\b_{kl}-\g_{kl}}{2}.
\ee
Once we plug in these final expressions in the definition for $\a'_{ij}$ (\ref{primeabc}), we can readily verify it is equal to the right hand side of (\ref{alphaprime}) (the $h^{'\dag}_i Z_1$ term is treated in the same manner as the identical one in $\b'_{ij}$), after we replace $\tilde c, {\bar{\tilde c}}$ from (\ref{c's}), and make use of the identity that follows from (\ref{b-c}),
\be\label{bg_identity}
(\b_{kl}-\g_{kl})(\b_{pq}-\g_{pq})=(\b_{kq}-\g_{kq})(\b_{pl}-\g_{pl}).
\ee

With equations \eqref{hprime}-\eqref{alphaprime} in place, expressing $Z_2$ in terms of the unprimed quantities becomes an almost trivial computation,
\be\ba
Z_2&=\alpha_{22}'Z_1+h_2'(h_2'^\dagger Z_1)\\
&=(\alpha_{22}-\frac{\alpha_{21}\alpha_{12}}{\alpha_{11}})Z_1-\frac{ c_{2,2}}{\alpha_{11} }(h_2^{'\dagger} Z_1)Z_1 +(h_2-\frac{\a_{12}}{\a_{11}}h_1)(h_2'^\dagger Z_1)+\frac{ c_{2,2}}{\alpha_{11} }(h_2^{'\dagger} Z_1)Z_1\\
&=(\alpha_{22}-\frac{\alpha_{21}\alpha_{12}}{\alpha_{11}})(\alpha_{11}+h_1 h_1^\dagger) Z_0 +(h_2-\frac{\a_{12}}{\a_{11}}h_1)(\alpha_{11}h_2^{\dagger}Z_{0}-\alpha_{21}h_1^{\dagger}Z_{0}),
\ea\ee
immediately yielding the final answer
\be
Z_2=(\a_{11}\a_{22}-\a_{12}\a_{21}+\a_{22}h_1 h^\dagger_1+\a_{11}h_2 h^\dagger_2-\a_{12}h_1 h^\dagger_2-\a_{21}h_2 h^\dagger_1)Z_0,
\ee
in agreement with \cite{Hatsuda:2009pc}. In particular notice that we never had to use the complicated form of $c_{2,2}$ since the terms containing it cancel out. The $N-$magnon derivation will heavily rely on the generalizations of \eqref{hprime}-\eqref{alphaprime}, which we will see in the next section.

\subsection{Building up recursion}\label{section_recursion}

Since the $N$-soliton solution is obtained recursively, we have to refine our notation in order to keep track of quantities corresponding to each intermediate dressed solution. Hence we can write for the solution of the spectral equation,
\be
\begin{aligned}
\Psi_N(\l)&=\chi_{N}(\l)\Psi_{N-1}(\l),\\
&=\chi_{N}(\l)\chi_{N-1}(\l)\ldots \chi_{1}(\l)\Psi_{0}(\l),
\end{aligned}
\ee
where $\Psi_{0}(\l)$ corresponds to the vacuum. As usual, $g_N=\Psi_N(0)$ and it can be expressed in terms of a projection operator
\be
g_N=\theta \left( 1-2P_N \right), \quad P_N=\frac{Z_N Z_N^\dagger}{|Z_N|^2}.
\ee

For the $CP^n$ coordinates $Z_N$ that span the projection operator, dressing amounts to
\be\label{Zn0}
Z_N=(\alpha^{(N-1)}_{NN}+h^{(N-1)}_N h^{(N-1)\dagger}_N) Z_{N-1},
\ee
where we define for all $N-1\ge m\ge 0$, $N\ge i,j\ge1$, the quantities\footnote{In fact, for our derivation we will only need the quantities with lower indices $N\ge i,j\ge m+1$.}
\be
h_i^{(m)}=\theta \Psi_m(\bar{\xi}_i) e_i,\quad\beta_{ij}^{(m)}=h_i^{(m)\dagger} h_j^{(m)}, \quad \gamma_{ij}^{(m)}=h_i^{(m)\dagger} \theta g_m h_j^{(m)},
\ee
and
\be
\alpha_{ij}^{(m)}=-\frac{\xi_i \beta_{ij}^{(m)}}{\xi_i-\bar{\xi}_j}-\frac{\gamma_{ij}^{(m)}}{\xi_i \bar{\xi}_j-1}.
\ee
In analogy with (\ref{Zn0}), the intermediate $(m+1)$-th step of dressing will be
\be\label{Zn}
Z_{m+1}=(\alpha^{(m)}_{m+1,m+1}+h^{(m)}_{m+1} h^{(m)\dagger}_{m+1}) Z_{m}.
\ee

For the case where the upper index in the last three equations becomes zero, we can drop it altogether and recover the notation of the previous sections.

Conversely, the recursive nature of the dressing method and the manner we have taken this into account in our definitions, allow us to obtain all relevant formulas for the $m\to m+1$ dressing from the $1\to2$ dressing of the previous section, simply by shifting all indices with specific, $0,1,2$ and so on, values\footnote{Remembering also that $h_i=h_i^{(0)}, h_i'=h_i^{(1)}$ and similarly for the $\a,\b,\g$ variables. The shift of indices is simply a consequence of the fact that every time we dress, $\xi_i\to\xi_{i+1}$ and $\Psi_{i-1}\to\Psi_{i}$, with all other quantities containing them being modified accordingly.}. For example, the generalization of \eqref{hiprime} will be
\be
h_j^{(m)}=h_j^{(m-1)}+c_{1,j}^{(m-1)}h_m^{(m-1)}+ c_{2,j}^{(m-1)}Z_{m-1}.
\ee

Most importantly, the straightforward generalization of relations \eqref{hprime}-\eqref{alphaprime} will be
\begin{align}
h_j^{(m)}&=h_j^{(m-1)}-\frac{\alpha_{mj}^{(m-1)}}{\alpha_{mm}^{(m-1)}}h_m^{(m-1)}+\frac{ c_{2,j}^{(m-1)}}{\alpha_{mm}^{(m-1)} }Z_m \label{id3}\\
\alpha_{ij}^{(m)}&=\alpha_{ij}^{(m-1)}-\frac{\alpha_{im}^{(m-1)}\alpha_{mj}^{(m-1)}}{\alpha_{mm}^{(m-1)}}-\frac{ c_{2,j}^{(m-1)}}{\alpha_{mm}^{(m-1)} }h_i^{(m)\dagger} Z_m\label{id2}\\
h_i^{(m)\dagger} Z_m&=\alpha_{mm}^{(m-1)}h_i^{(m-1)\dagger}Z_{m-1}-\alpha_{im}^{(m-1)}h_m^{(m-1)\dagger}Z_{m-1}\;. \label{id1}
\end{align}
Notice that we can replace the last term on the right of \eqref{id3} with \eqref{Zn}, and similarly with \eqref{id2} and \eqref{id1}, so that we only have quantities of the $(m-1)$-dressed solution, if desired.

The derivation and proof of the $N$-magnon solution will rely solely on the use of the above relations, together with \eqref{Zn}. As it was the case for $N=2$, we will not be needing the form of the $c_{2,j}$ coefficient, since we'll show that it drops out of the calculation.

\section{$N$-soliton solutions}

\subsection{$CP^n$ case}

Let us start by stating our main result. The $CP^n$ coordinate vector for the $N$-magnon solution will be given by
\be\label{Zn=}
Z_N=\left( {\rm det} (\a_{ij})+\sum_{i,j=1}^N (-1)^{i+j} M_{ij}\, h_j h_i^\dagger \right) Z_0,
\ee
where $M_{ij}$ is the minor formed by removing the $i$'th row and $j$'th column from the matrix with elements $\a_{ij}$, $i,j=1\ldots N$.

The key step that unveils the structure of the $N$-magnon solution and facilitates its proof, is to realize that we can write it in the form
\be
Z_N=\left|
\begin{array}{cccc}
Z_{0}&-h_{1}&\ldots& -h_N\\
h_{1}^\dagger Z_{0}&\alpha_{11}&\ldots  & \alpha_{1N}\\
\vdots&\ldots&&\vdots\\
h_{N}^\dagger Z_{0}&\alpha_{N1}&\ldots  & \alpha_{NN}
\end{array}
\right|\;,\label{Znfinal}
\ee
where the elements of the first row are the only ones that are vectors instead of scalars. This is not a problem however, since we can analyze all vectors in a common basis and consider the above relation for each of the components.

In order to proceed with the proof, we will first need to define the following set of determinants,
\be\label{A(nk)}
A^{(N,k)}=\left|
\begin{array}{cccc}
Z_{N-k}&-h_{N-k+1}^{(N-k)}&\ldots& -h_N^{(N-k)}\\
h_{N-k+1}^{(N-k)\dagger}Z_{N-k}&\alpha_{N-k+1,N-k+1}^{(N-k)}&\ldots  & \alpha_{N-k+1,N}^{(N-k)}\\
\vdots&\ldots&&\vdots\\
h_{N}^{(N-k)\dagger}Z_{N-k}&\alpha_{N,N-k+1}^{(N-k)}&\ldots  & \alpha_{NN}^{(N-k)}
\end{array}
\right|\;,
\ee
corresponding to $k+1$-dimensional matrices, with $0\le k\le N$.

As a first step towards proving the $N$-magnon formula, we will now prove that
\be\label{A^k_to_A^k+1}
A^{(N,k)}=A^{(N,k+1)},
\ee
which obviously implies $A^{(N,k)}=A^{(N,m)}$ for any $k,m$ and hence provides us with a large set of identities between the quantities corresponding to intermediate dressed solutions.

The proof is solely based on elementary row and column operations that leave a determinant invariant. We start with \eqref{A(nk)} and use the identities \eqref{id3}-\eqref{id2} aiming to express $A^{(N,k)}$ in terms of the quantities of the $(N-k-1)$-dressed solution. In all columns apart from the first one, the terms multiplying $c_{2,j}^{(N-k-1)}$ will be proportional to the first column. Hence we can eliminate these terms simply by adding the first column to the remaining ones, multiplied by a suitable factor.

After performing these simplifications, and apply the replacements (\ref{Zn}) and (\ref{id1}), we see that $A^{(N,k)}$ can equivalently be given as the determinant of the matrix with the following matrix elements,
\be\label{calA(n,k)}
{\cal A}^{(N,k)}_{ij}=\begin{cases}
\alpha^{(N-k-1)}_{N-k,N-k}Z_{N-k-1}+h^{(N-k-1)}_{N-k} h^{(N-k-1)\dagger}_{N-k} Z_{N-k-1}&i=j=0\\
\\
-h_{N-k+j}^{(N-k-1)}+\frac{\alpha_{N-k,N-k+j}^{(N-k-1)}}{\alpha_{N-k,N-k}^{(N-k-1)}}h_{N-k}^{(N-k-1)}&i=0,\;j=1,\ldots k\\
\\
\alpha_{N-k,N-k}^{(N-k-1)}\;h_{N-k+i}^{(N-k-1)\dagger}Z_{N-k-1}-\alpha_{N-k+i,N-k}^{(N-k-1)}\;h_{N-k}^{(N-k-1)\dagger}Z_{N-k-1}&i=1,\ldots k,\;j=0\\
\\
\alpha_{N-k+i,N-k+j}^{(N-k-1)}-\frac{\alpha_{N-k+i,N-k}^{(N-k-1)}\alpha_{N-k,N-k+j}^{(N-k-1)}}{\alpha_{N-k,N-k}^{(N-k-1)}}&i,j=1,\ldots k,\;.
\end{cases}
\ee

Once we have massaged $A^{(N,k)}$ in this way, we turn to $A^{(N,k+1)}$, as defined in (\ref{A(nk)}), and
\begin{enumerate}
\item Multiply the first column with $\alpha_{N-k,N-k}^{(N-k-1)}$.
\item Divide the second row with $\alpha_{N-k,N-k}^{(N-k-1)}$.
\item Multiply the second row with $h_{N-k}^{(N-k)}$ and add it to the first row.
\item Similarly, for $m=2,\ldots,k+1$, multiply the second row with $-\alpha_{N-k-1+m,N-k}^{(N-k-1)}$ and add it to the $(m+1)$-th row.
\item Exchange the first and second rows.
\item Exchange the first and second columns.
\end{enumerate}
In this manner, we end up with
\be\label{A(nk+1)}
A^{(N,k+1)}=\left|
\begin{array}{ccccc}
1&h_{N-k}^{(N-k-1)\dagger}Z_{N-k-1}&\frac{\alpha_{N-k,N-k+1}^{(N-k-1)}}{\alpha_{N-k,N-k}^{(N-k-1)}}&\ldots  & \frac{\alpha_{N-k,N}^{(N-k-1)}}{\alpha_{N-k,N-k}^{(N-k-1)}}\\
0&&&&\\
\vdots&&{\cal A}^{(N,k)}&&\\
0&&&&
\end{array}
\right|\;,
\ee
where we mean that the submatrix in the lower right corner takes the form (\ref{calA(n,k)}). Now clearly $A^{(N,k+1)}=\det {\cal A}^{(N,k)}=A^{(N,k)}$, and hence we have proven the intermediate step (\ref{A^k_to_A^k+1}).

Then, (\ref{Znfinal}) clearly follows from the equality between the particular determinants $A^{(N,0)}=Z_{N}$ and $A^{(N,N)}$, thus completing the proof of the $N$-soliton solution.

%======================================================================================================================
\subsection{$SU(n)$ case}\label{subsection_SU(n)}

It is natural to ask whether our method of obtaining general multisoliton solutions by solving the recursion of the dressing method, could be applied to different symmetric spaces as well. The next obvious choice to consider would be the $SU(n)$ sigma model, for which the dressing takes a particularly simple form \cite{Zakharov:1980ty,Harnad:1983we,Spradlin:2006wk,Hollowood:2009tw}, and moreover the $n=2$ case is directly relevant to the discussion of giant magnons.

The $SU(2)$ sigma model is equivalent to string theory on the $R\times S^3$ subspace of $AdS_5 \times S^5$ and by dressing one obtains scattering states of dyonic giant magnons \cite{Chen:2006gea} on this background. The $N$-soliton solutions we will derive here are precisely of this form for $n=2$, providing generalizations, for arbitrary vacuum and soliton polarizations, of our earlier work \cite{Kalousios:2008gz}, and complementing the approach of \cite{Harnad:1982cf}.

For simplicity, we will perform the calculation for the $U(n)$ sigma model, and point out the very simple modifications required for obtaining the $SU(n)$ result at the very end. For $U(n)$ dressing, the auxiliary field
obeys only one constraint, that of unitarity,
\be\label{constraint44}
[\Psi(\bar{\l})]^\dagger \Psi(\l)=1,
\ee
and the simplest dressing factor can have one pole \cite{Harnad:1983we}. In notation adapted from \cite{Spradlin:2006wk} to better suit recursion, if we start from the vacuum $\Psi_0$ and dress consecutively, at the $N$-th step we obtain $\Psi_N(\l)=\chi_N(\l)\Psi_{N-1}(\l)$, with a dressing factor
\be\label{chi_N}
\chi_N(\l)=1+\frac{\xi_N-\bar{\xi}_N}{\l-\xi_N}{\cal P}_N.
\ee
The rank one hermitian projection operator ${\cal P}_N$ may be expressed as
\be
{\cal P}_N=\frac{1}{\b_{NN}^{(N-1)}}\; h_N^{(N-1)} h_N^{(N-1)\dagger},
\ee
where we define
\be
h_i^{(N)}=\Psi_N(\bar{\xi}_i)e_i, \quad \b_{ij}^{(N)}=h_i^{(N)\dagger} h_j^{(N)},\quad \a_{ij}^{(N)}=-\frac{\xi_i \b_{ij}^{(N)}}{\xi_i-\bar{\xi_j}},
\ee
for any $N\ge0$. Then, the matrix field $g_N=\Psi_N(0)$ will be given by
\be\label{gn}
g_N=\frac{1}{\alpha^{(N-1)}_{NN}}\big(\alpha^{(N-1)}_{NN}+h^{(N-1)}_N h^{(N-1)\dagger}_N\big) g_{N-1},
\ee
and with the help of (\ref{chi_N})-(\ref{gn}), we can similarly obtain recursion relations for the following quantities of interest,
\begin{align}
h_i^{(N)} &= h_i^{(N-1)}- \frac{\a_{Ni}^{(N-1)}}{\a_{NN}^{(N-1)}} h_N^{(N-1)} \label{hSU}\\
\a_{ij}^{(N)} &=\a_{ij}^{(N-1)}-\frac{\a_{iN}^{(N-1)}\a_{Nj}^{(N-1)}}{ \a_{NN}^{(N-1)}} \label{aSU}\\
h_i^{(N)\dagger} g_N &= \frac{ 1}{\a_{NN}^{(N-1)}}\big(\a_{NN}^{(N-1)} h_i^{(N-1)\dagger}g_{N-1} -\a_{iN}^{(N-1)} h_N^{(N-1)\dagger}g_{N-1}\big).\label{hgSU}
\end{align}

From the last four relations, it is possible to solve the recursion for $g_N$, in order to obtain\be\label{gn=}
g_N=\frac{1}{{\rm det} (\a_{ij})}\left( {\rm det} (\a_{ij})+\sum_{i,j=1}^N (-1)^{i+j} M_{ij}\, h_j h_i^\dagger \right) g_0,
\ee
where $M_{ij}$ is the minor formed by removing the $i$'th row and $j$'th column from the matrix with elements $\a_{ij}$, $i,j=1\ldots N$. Equivalently, we may write this as
\be\label{gnDet}
g_N=\frac{1}{{\rm det} (\a_{ij})}
:\begin{vmatrix}
 g_0 & -h_1 & \cdots & -h_N \\
h_1^\dagger g_0 & \a_{11} & \cdots & \a_{1N} \\
\vdots & \vdots & \vdots & \vdots \\
h_N^\dagger g_0 & \a_{N1} & \cdots & \a_{NN}
\end{vmatrix}:~,
\ee
where the colons around the second determinant simply mean that upon expanding the determinant, the column $h_i$ is ordered before the row $h_j^\dagger g_0$  \footnote{We can avoid the need for an ordering notation by breaking up (\ref{gnDet}) into $n$ relations, one for each column of the matrices $g_i$. This is possible because (\ref{gn}) and (\ref{hgSU}) hold for each column separately, and hence the notation reduces to that of the $CP^n$ case.\label{footnote_SU(n)}}.

The proof of (\ref{gn=})-(\ref{gnDet}) is a straightforward generalization of the $CP^n$ proof.  Namely, we start by defining the following set of determinant ratios,
\be
C^{(N,k)}=\frac{A^{(N,k)}}{B^{(N,k)}},
\ee
where $A^{(N,k)}$ is just like (\ref{A(nk)}) with $Z\to g$, together with the proper ordering,
\be\label{A(nk)2}
A^{(N,k)}=\;
:\left|
\begin{array}{cccc}
g_{N-k}&-h_{N-k+1}^{(N-k)}&\ldots& -h_N^{(N-k)}\\
h_{N-k+1}^{(N-k)\dagger}g_{N-k}&\alpha_{N-k+1,N-k+1}^{(N-k)}&\ldots  & \alpha_{N-k+1,N}^{(N-k)}\\
\vdots&\vdots&\vdots&\vdots\\
h_{N}^{(N-k)\dagger}g_{N-k}&\alpha_{N,N-k+1}^{(N-k)}&\ldots  & \alpha_{N,N}^{(N-k)}
\end{array}
\right|:~,
\ee
and $B^{(N,k)}$ can be obtained from $A^{(N,k)}$ by removing the first row and column,
\be\label{B(nk)2}
B^{(N,k)}=
\begin{vmatrix}
\alpha_{N-k+1,N-k+1}^{(N-k)}&\ldots  & \alpha_{N-k+1,N}^{(N-k)}\\
\vdots&\vdots&\vdots\\
\alpha_{N,N-k+1}^{(N-k)}&\ldots  & \alpha_{N,N}^{(N-k)}
\end{vmatrix}~,
\ee
with $0\le k\le N$. Proving the relation between the $A$'s for consecutive $k$ is almost identical to the $CP^n$ case, the only difference being an extra $1/\a$ factor in the first column of $A^{(N,k)}$ due to (\ref{gn}) and (\ref{hgSU}). Namely,
\be\label{Ank=}
A^{(N,k)}= \frac{A^{(N,k+1)}}{\a^{(N-k-1)}_{N-k,N-k}}.
\ee
In a similar fashion, we may also prove that
\be\label{Bnk=}
B^{(N,k)}= \frac{B^{(N,k+1)}}{\a^{(N-k-1)}_{N-k,N-k}}.
\ee
In more detail, we start from the right-hand side of (\ref{Bnk=}) and
\begin{enumerate}
 \item absorb the $1/\a_{N-k,N-k}^{(N-k-1)}$ factor in the first row of $B^{(N,k+1)}$
 \item for $m=1,\ldots,k$ multiply the first row with $-\a_{N-k+m,N-k}^{(N-k-1)}$ and add this to the $(m+1)$-th row,
\end{enumerate}
which precisely gives $B^{(N,k)}$, after we perform the replacement (\ref{aSU}) in (\ref{B(nk)2}). Clearly, from the above relations we have that
\be
C^{(N,k)}= C^{(N,k+1)},
\ee
which also implies that $C^{(N,0)}=g_N$ and $C^{(N,N)}$ will be equal, hence proving (\ref{gnDet}).

Finally, we need to describe the transition from $U(n)$ to $SU(n)$. For a matrix field $\tilde g_N$ to belong to $SU(n)$, the additional constraint $\det \tilde g_N=1$ needs to be satisfied. Since \eqref{chi_N} implies\footnote{Proving this relation requires utilizing the determinant-to-trace relation and the projection operator properties ${\rm Tr}[{\cal P}_N]=1$ and ${\cal P}_N^j={\cal P}_N$ for any positive integer $j$.}
\be
\det \chi_N (\l)=\frac{\l-\bar\xi_N}{\l- \xi_N},
\ee
it is evident that we have to rescale the dressing phase as
\be
\tilde \chi_N(\l)=\left(\frac{\xi_N}{\bar\xi_N}\right)^{1/n}\chi_N(\l)
\ee
in order satisfy the constraint. Hence for an $SU(n)$ vacuum $\tilde g_0=g_0$ we have
\be
\ba
\tilde g_N&=\tilde \chi_N(0)\tilde\chi_{N-1}(0)\ldots\tilde\chi_{1}(0)\tilde g_0\\
&=\prod_{i=1}^N \left(\frac{\xi_i}{\bar\xi_i}\right)^{1/n} \chi_N(0)\chi_{N-1}(0)\ldots\chi_{1}(0) g_0\\
&=\prod_{i=1}^N \left(\frac{\xi_i}{\bar\xi_i}\right)^{1/n}g_N,
\ea
\ee
from which we infer that the matrix field $\tilde g_N$, representing the $N$-soliton solution in $SU(n)$, will be given by
\be
\tilde g_N=\frac{\prod_{i=1}^N \left(\frac{\xi_i}{\bar\xi_i}\right)^{1/n}}{{\rm det} (\a_{ij})}\left( {\rm det} (\a_{ij})+\sum_{i,j=1}^N (-1)^{i+j} M_{ij}\; h_j h_i^\dagger \right) \tilde g_0,
\ee
where $M_{ij}$ is the minor formed by removing the $i$'th row and $j$'th column from the matrix with elements $\a_{ij}$, $i,j=1\ldots N$.

%========================================================
\subsection{$S^n$ case}

Since a main motivation for this work has been giant magnons in the $AdS_5/CFT_4$ and $AdS_4/CFT_3$ correspondence, if we would like to obtain multimagnon solutions that extend beyond the $R\times S^3$ subspace of $AdS_5 \times S^5$, we would have to consider $S^n = SO(n+1)/SO(n)$ dressing. We briefly examine how our method works for this particular symmetric space in the current section.

In $S^n$, there is just one additional constraint on the auxiliary function $\Psi$ compared to the $CP^n$ case \cite{Harnad:1983we}, which is suggestive of the relation between the two symmetric spaces. Altogether we have
\be\label{Sn_constraints}
[\Psi(\bar{\l})]^\dagger \Psi(\l)=1, \qquad \Psi(\l)=\Psi(0)\theta \Psi(1/\l) \theta, \qquad \overline{\Psi(\bar{\l})}=\Psi(\l),
\ee
and the above constraints have two distinct `minimal' solutions for the dressing factor. The simplest one has two simple poles
\be\label{twopoles}
\xi_1, \bar{\xi}_1 = 1/ \xi_1,
\ee
and gives rise to Hofman-Maldacena giant magnons \cite{Hofman:2006xt}, whereas the second one has four poles,
\be
\xi_1,\bar{\xi}_1, 1/\xi_1 ,1 / \bar{\xi}_1
\ee
yielding dyonic giant magnons \cite{Chen:2006gea}, as was explained in \cite{Hollowood:2009sc}. In what follows, we will only focus on the first case (\ref{twopoles}), where our knowledge of the $CP^n$ dressing will be most useful.

The corresponding dressing factor is given by \cite{Saint Aubin:1982an,Spradlin:2006wk}
\be\label{Sn_chi1}
\chi_1(\l)=1+\frac{\xi_1-\bar{\xi}_1}{\l-\xi_1} \mathcal{P}_1 +\frac{\bar{\xi}_1 -\xi_1}{\l-\bar{\xi}_1} \overline{\mathcal{P}}_1,
\ee
where the projector $\mathcal{P}_1$ is given by
\be
\mathcal{P}_1=\frac{\Psi_0(\bar{\xi}_1)e_1 e_1^\dagger \Psi_0^{-1}(\xi_1)} {e_1^\dagger \Psi_0^{-1}(\xi_1) \Psi_0(\bar{\xi}_1)e_1},
\ee
with (\ref{Sn_constraints}) implying two additional conditions for the polarization vectors $e_i$,
\be\label{e1_constraints}
\bar e_i=\theta e_i,\quad e_i^T e_i=0.
\ee
Using (\ref{Sn_constraints}), (\ref{twopoles}) and (\ref{e1_constraints}), it is possible to recast (\ref{Sn_chi1}) in the more familiar form
\be
\chi_1(\l) = 1+\frac{Q_1}{\l -\xi_1} + \frac{Q_2}{\l-1/\xi_1},
\ee
where
\be
Q_1 = \frac{1}{\Delta} \left(-\frac{\b_{11}}{\xi_1 -1/\xi_1} \theta h_1 h_1^\dagger \theta \right) , \qquad Q_2 = \frac{1}{\Delta} \left(\frac{\b_{11}}{\xi_1-1/\xi_1} g_0 h_1 h_1^\dagger g_0^\dagger \right)
\ee
and also
\be\label{sh=}
h_i=\theta \Psi_0(\bar{\xi}_i) e_i, \qquad \b_{ij}= h_i^\dagger h_j, \qquad \Delta = - \frac{\b_{11}^2}{(\xi_1-1/\xi_1)^2}.
\ee
Comparing with (\ref{chi_1})-(\ref{Delta}), we clearly see that the $S^n$ dressing factor may be obtained from the $CP^n$ one by taking, in this particular order,
\be
\gamma_{ii}\to0\qquad\bar\xi_i\to1/\xi_i.
\ee\label{Sn_replacement}
Indeed, if we define here as well
\be\label{sgamma=}
\g_{ij}=h_i^\dagger \theta g_0 h_j,
\ee
then is is easy to show that it vanishes for $i=j$,
\be\label{gamma=0_proof}
\g_{ii}=e_i^\dagger \Psi_0^\dagger(\bar\xi_i)\theta \theta g_0 \theta \Psi_0(\xi_i)e_i=e_i^\dagger \Psi_0^\dagger(1/\xi_i)\Psi_0(1/\xi_i)\theta e_i=e_i^T e_i=0,
\ee
where we have again made use of (\ref{Sn_constraints}), (\ref{twopoles}) and (\ref{e1_constraints}).

Hence, we arrive at the important conclusion that we may obtain all $S^n$ dressing formulas from their $CP^n$ analogs, with the help of the replacement (\ref{Sn_replacement}). For example, the definition for the $\a$ variables (\ref{cpalpha}) will now give,
\be\label{salpha=}
\a_{ij}=
\begin{cases}
-\frac{\xi_i \beta_{ij}}{\xi_i-1/\xi_j}-\frac{\gamma_{ij}}{\xi_i /\xi_j-1} & i \neq j \\
\\
-\frac{\xi_i \beta_{ii}}{\xi_i-1/\xi_i} & i=j,
\end{cases}
\ee
which also implies that $\Delta$ may be expressed as
\be
\Delta=\a_{11}\bar\a_{11}.
\ee
More importantly, the $N$-th dressed matrix fields $g_N$ will be obtained in the same fashion\footnote{In the generalized notation of Section \ref{section_recursion}, we can also prove that $\gamma^{(N)}_{ii}=0$, similarly to (\ref{gamma=0_proof}).}, as well as the corresponding projection operators $P_N=\frac 1 2 (1-\theta g_N)$. The only minor subtlety is that from the $CP^n$ case we only know the complex eigenvectors of $P_N$, with arbitrary length,
\be
P_N=\frac{Z_N Z_N^\dagger}{|Z_N|^2}\Rightarrow P_N Z_N=Z_N,
\ee
whereas for $S^n$, the target space vectors should be real, with norm equal to one,
\be\label{Sn_Pn}
P_N=X_N X_N^T \Rightarrow P_N X_N=X_N,\;\text{with $\bar X_N=X_N, |X_N|=1$}.
\ee
Let us show how to obtain $X_N$ from $Z_N$, focusing for clarity on the single magnon case. We start by stating a number of important identities, which may be straightforwardly derived from the constraints (\ref{Sn_constraints}), (\ref{twopoles}) and (\ref{e1_constraints}). In particular, we can show that
\be
\bar h_i=\theta g_0 h_i,
\ee
that all $\beta_{ij}$ and $\gamma_{ij}$ are real and symmetric,
\be
\ba
\beta_{ij}=\bar\beta_{ij}=\beta_{ji}\\
\gamma_{ij}=\bar\gamma_{ij}=\gamma_{ji},
\ea
\ee
and finally, that
\be
a_{ij}+\bar a_{ij}=-(\beta_{ij}-\gamma_{ij}).
\ee
Then, starting with an $S^n$ vacuum $Z_0=X_0$, $\bar Z_0=Z_0$ and $|Z_0|=1$, we can prove that $Z_1$ is in fact imaginary,
\be\ba
\bar Z_1&=(\bar \a_{11}+\bar h_1 \bar h_1^\dagger)Z_0\\
&=\bar \a_{11}Z_0+(1-2P_0)h_1 h_1^\dagger(1-2P_0)Z_0\\
&=\bar \a_{11}Z_0-h_1 h_1^\dagger-(\a_{11}+\bar \a_{11})Z_0\\
&=-(\a_{11}+h_1 h_1^\dagger)Z_0=-Z_1,
\ea\ee
from which it follows that if we define
\be
X_1=\frac{-i Z_1}{|Z_1|},
\ee
this clearly satisfies (\ref{Sn_Pn}) and hence gives the correct $S^n$ coordinate vector for a single magnon. By the same token, if $Z_1$ is imaginary, then $Z_2$ will be real and so forth, so that the $N$-magnon vector for the $S^n$ sigma model will be given by
\be\label{Xn=}
X_N=\frac{(-i)^N Z_N}{|Z_N|},
\ee
where $Z_N$ is given by (\ref{Znfinal}), with the appropriate $S^n$ definitions (\ref{sh=}),(\ref{sgamma=}) and (\ref{salpha=}).

\section{Classical Time Delay}

Let us demonstrate the utility of our solutions by calculating the classical time delay \cite{Jackiw:1975im} resulting from $N$-soliton scattering. We will be examining the case where the constituents are the $CP^2$ elementary dyonic magnons, which were first discovered in \cite{Abbott:2009um} for specific values of the soliton parameters, and generalized to all possible kinematic configurations in \cite{Hollowood:2009sc}.

The form of the $N$-soliton solution for the specific constituents has been derived in appendix \ref{appx_CP3}, and the time delay will be inferred by looking at a particular limit of large magnon separation when $t\to\pm\infty$, in the lines of \cite{1972JPSJ...33.1459H, Kalousios:2008gz}. To this end, we define the real quantities
\be\label{uw}
\begin{aligned}
u_l&\equiv i (\mathcal{Z}_l-\bar{\mathcal{Z}}_l)=
\kappa_l x- \nu_l t,\\
w_l&\equiv \mathcal{Z}_l+\bar{\mathcal{Z}}_l,\\
\end{aligned}
\ee
with
\be
\begin{aligned}
\kappa_l&=-2i\frac{ (\xi_l-\bar\xi_l)(1+|\xi_l|^2)}{\left|1-\xi_l\right|^2\left|1+\xi_l\right|^2},\\
\nu_l&=\frac{-2i(\xi_l^2-\bar\xi_l^2)}{\left|1-\xi_l\right|^2\left|1+\xi_l\right|^2},
\end{aligned}
\ee
according to (\ref{calZ_in_CP}). The parameter $u_l$ controls the soliton profiles, as it can be seen from the exponential factors in (\ref{Z_Nminus})-(\ref{Z_1}) and the fact that
\be
i  \mathcal{Z}_l=\frac{1}{2}(u_l+ i w_l),~~~-i \bar{\mathcal{Z}}_l=\frac{1}{2}(u_l- i w_l).
\ee
Without loss of generality, we may assume that the magnons are ordered such that their velocities obey
\be
\frac{\nu_1}{\kappa_1}>\frac{\nu_2}{\kappa_2}>...>\frac{\nu_N}{\kappa_N},
\ee
and restrict the soliton parameters $\xi_i$ such that all $\kappa_l$ are positive.

We now zoom into the $k$-th magnon as $t\to\pm\infty$. This means that $u_k$ should remain finite in that limit, implying that $x$ should scale as $x=\frac{\nu_k}{\kappa_k} t+\frac{u_k}{\kappa_k}$. So, in total the
$u_l$ will behave as
\be
u_l=\kappa_l \left(\frac{\nu_k}{\kappa_k}-\frac{\nu_l}{\kappa_l}\right)t+\kappa_l\frac{u_k}{\kappa_k},
\ee
and the limit $t\to-\infty$ under the aforementioned ordering and scaling implies
\bea
&&u_1, u_2,\ldots,u_{k-1}\to+\infty\nonumber,\\
&&u_k ~~\text{finite},\\
&&u_{k+1},u_{k+2},\ldots,u_N\to-\infty\nonumber,
\eea
whereas for $t\to+\infty$ we just exchange $+\infty\leftrightarrow -\infty$.

Let us first focus in the $t\to-\infty$ limit, and start from the $Z_N^-$ component. The main trick is to remove a factor $e^{i \mathcal{Z}_l}$ from the $l$-th row, and a factor $e^{-i \mathcal{\bar Z}_l}$ for the $l$-th column, $l=1,\ldots, k-1$, since they are the divergent terms. This way, all the divergences are gathered in the overall factor
\be\label{divergent_factor}
\prod_{i=1}^{k-1} e^{i(\mathcal{Z}_i-\mathcal{\bar Z}_i)}=\prod_{i=1}^{k-1}e^{u_i},
\ee
which we can drop because of the equivalence $Z\sim \mu Z$, leaving what remains in the determinant finite. In particular, the finite part will be
\be
Z_N^-=e^{it}\left|
\begin{array}{ccc|c|ccc}
&&&&&&\\
&\frac{-2\xi_i}{\xi_i-\bar\xi_j}&  &\frac{-2\xi_i}{\xi_i-\bar\xi_k}e^{-i \mathcal{\bar Z}_k}&&\bf{0}&\\
&&&&&&\\\hline
&\frac{-2\xi_k}{\xi_k-\bar\xi_j}e^{i \mathcal{Z}_k}&  &\alpha_{kk}&&K_{kj}&\\\hline
&&&&&&\\
&\bf{0}& &K_{ik}&\quad&K_{ij}&\quad\\
&&&&&&\\
\end{array}\right|,
\ee
where in the upper left and lower right corner we have square $(k-1)\times(k-1)$ and $(N-k)\times(N-k)$ submatrices respectively, in the upper right and lower left corner we have $(k-1)\times(N-k)$ and $(N-k)\times(k-1)$ zero submatrices respectively, and the $k$-th row and column are singled out.

If we now expand the determinant with respect to the $k$-th column (or row), we notice that all submatrices multiplying the column (or row) elements are of block triangular form, and we can use the identities
\be
\left|
\begin{array}{cccc}
A& 0\\
C& D
\end{array}\right|=
\left|
\begin{array}{cccc}
A& B\\
0& D
\end{array}\right|=|A||D|
\ee
where $A$ and $D$ are square submatrices, in order to simplify the calculation. In this fashion, we obtain for the finite part of $Z_N^-$,
\be\label{Z_minus_at_-inf}
Z_N^-=\prod_{i=1}^{k-1}(-2\xi_i)\; \Xi^{1,k-1}\;\frac{\mathcal{K}^{k,N}}{K_{kk}}\;e^{it}\left(e^{u_k+\delta u_k^-}\frac{-2\xi_k}{\xi_k-\bar\xi_k}+K_{kk}\right),
\ee
where
\be
\begin{aligned}
\Xi^{l,m}&=\det(\frac{1}{\xi_i-\bar\xi_j}),&\\
{\cal K}^{l,m}&=\det (K_{ij}),&\text{with}\quad i,j=l,l+1,\ldots,m\;,\\
K_{ij}&=\frac{(1-\xi_i^2)\bar \xi_j {\bf\Omega}^\dagger_i {\bf \Omega}_j}{(\xi_i-\bar \xi_j)(\xi_i\bar \xi_j-1)},&
\end{aligned}
\ee
 with ${\bf\Omega}_i$ a 2-dimensional component vector of $e_i$, defined through $e_i=(1~~ i ~~ {\bf \Omega}_i)^{\rm T}$ (see appendix \ref{appx_CP3} for more details). Finally, in (\ref{Z_minus_at_-inf}) we also have
\be
e^{\delta u_k^-}=\prod_{i=1}^{k-1}\left|\frac{\xi_i-\xi_k}{\bar\xi_i-\xi_k}\right|^2\frac{\mathcal{K}^{k+1,N}}{\mathcal{K}^{k,N}}K_{kk}.
\ee
The calculation is almost identical for $Z_N^+$, and only slightly different for $Z_N^\perp$, yielding
\be\label{Z+Zperp}
\begin{aligned}
Z_N^+&=\prod_{i=1}^{k-1}(-2\bar\xi_i)\; \Xi^{1,k-1}\;\frac{\mathcal{K}^{k,N}}{K_{kk}}\;e^{-it}\left(e^{u_k+\delta u_k^-}\frac{-2\bar\xi_k}{\xi_k-\bar\xi_k}+K_{kk}\right),\\
Z_N^\perp&=\prod_{i=1}^{k-1}(-2\bar\xi_i)\;\Xi^{1,k-1}\prod_{i=1}^{k-1}\frac{\xi_i-\xi_k}{\bar\xi_i-\xi_k}\left(-e^{-it}e^{i \mathcal{Z}_k}{\bf \Omega}^{k,N}_\mathcal{K}\right),
\end{aligned}
\ee
where ${\bf \Omega}^{k,N}_\mathcal{K}$ is the vector that arises if we replace $K_{ki}\to{\bf \Omega}_i$ in the $\mathcal{K}^{k,N}$ determinant, for all $i$.

The same procedure can be repeated for the $t\to+\infty$ limit, this time removing the divergent $e^{i \mathcal{Z}_l}$ and $e^{-i \mathcal{\bar Z}_l}$ exponentials from rows and columns $l=k+1,\ldots, N$ respectively. We obtain
\be\label{Z_at_t_to_+inf}
\begin{aligned}
Z_N^-&=\prod_{i=k+1}^{N}(-2\xi_i)\; \Xi^{k+1,N}\;\frac{\mathcal{K}^{1,k}}{K_{kk}}\;e^{it}\left(e^{u_k+\delta u_k^+}\frac{-2\xi_k}{\xi_k-\bar\xi_k}+K_{kk}\right),\\
Z_N^+&=\prod_{i=k+1}^{N}(-2\bar\xi_i)\; \Xi^{k+1,N}\;\frac{\mathcal{K}^{1,k}}{K_{kk}}\;e^{-it}\left(e^{u_k+\delta u_k^+}\frac{-2\bar\xi_k}{\xi_k-\bar\xi_k}+K_{kk}\right),\\
Z_N^\perp&=\prod_{i=k+1}^{N}(-2\bar\xi_i)\;\Xi^{k+1,N}\prod_{i=k+1}^{N}\frac{\xi_i-\xi_k}{\bar\xi_i-\xi_k}\left(-e^{-it}e^{i \mathcal{Z}_k}{\bf \Omega}^{1,k}_\mathcal{K}\right),
\end{aligned}
\ee
where
\be
e^{\delta u_k^+}=\prod_{i=k+1}^{N}\left|\frac{\xi_i-\xi_k}{\bar\xi_i-\xi_k}\right|^2\frac{\mathcal{K}^{1,k-1}}{\mathcal{K}^{1,k}}K_{kk},
\ee
and similarly with before, ${\bf \Omega}^{1,k}_\mathcal{K}$ is the vector that arises if we replace $K_{ki}\to{\bf \Omega}_i$ in the $\mathcal{K}^{1,k}$ determinant, for all $i$.

As pointed out in \cite{Hollowood:2009tw}, changing the norm of the individual soliton polarizations ${\bf \Omega}_i$ simply causes shifts in $x$, so for simplicity we can set all $|{\bf \Omega}_i|=1$. Clearly the $Z_N^{\pm}$ components in (\ref{Z_minus_at_-inf}), (\ref{Z+Zperp}) and (\ref{Z_at_t_to_+inf}) have the form of the 1-magnon solution (\ref{Z_1}), up to rescalings and overall phases that can be different for each component \cite{Kalousios:2008gz}. One can verify that the seemingly different $Z_N^\perp$ components also have the desired form, up to an allowed overall phase, by noticing that
\be
|{\bf \Omega}^{k,N}_\mathcal{K}|^2=\frac{\mathcal{\bar K}^{k+1,N}\mathcal{K}^{k,N}}{K_{kk}}\;,\quad\quad|{\bf \Omega}^{1,k}_\mathcal{K}|^2=\frac{\mathcal{\bar K}^{1,k-1}\mathcal{K}^{1,k}}{K_{kk}}.
\ee

Hence we have proven that in the limit of large magnon separation, our $N$-magnon solution breaks up into a linear superposition of single magnons, as expected from integrability. What is very interesting however, is that the interaction changes the soliton polarizations, as it can be seen by comparing (\ref{Z+Zperp}) and (\ref{Z_at_t_to_+inf}). This phenomenon was first accounted in the vector nonlinear Schroedinger equation \cite{Manakov}, and is a generic feature of solitons that are characterized by a polarization vector of dimension more than one\footnote{We thank A.~V.~Mikhailov for pointing this out to us.}.

From our results, we can infer that the shift in the parameter $u_k$ due to magnon scattering is $\delta u_k=\delta u_k^+-\delta u_k^-$, thus
\be\label{u_shift}
e^{\delta u_k}=\prod_{i=1}^{k-1}\left|\frac{\bar\xi_i-\xi_k}{\xi_i-\xi_k}\right|^2\prod_{i=k+1}^{N}\left|\frac{\xi_i-\xi_k}{\bar\xi_i-\xi_k}\right|^2\frac{{\cal K}^{1,k-1}}{{\cal K}^{1,k}}\frac{{\cal K}^{k,N}}{{\cal K}^{k+1,N}},
\ee
and due to (\ref{uw}), the corresponding time delay will be given by
\be\label{time_delay}
\delta t_k=\frac{\delta u_k}{\nu_k}=2i\frac{\left|1-\xi_k\right|^2\left|1+\xi_k\right|^2}{(\xi_k^2-\bar\xi_k^2)}\;\delta u_k.
\ee
One can immediately check that for $N=2, k=1$, our results (\ref{u_shift})-(\ref{time_delay}) agree with the time delay computation of \cite{Hatsuda:2009pc}, up to an overall factor of $2$, owing to the different convention $t\leftrightarrow 2t$.

Furthermore, if we choose all magnon polarizations to lie in either of two orthogonal axes, for example ${\bf \Omega}_i=(1~~0)^T$ or ${\bf \Omega}_i=(0~~1)^T$ as in the latter reference, we can show that
\be
\delta t_k=-\sum_{i=1}^{k-1}\delta t_{k,i}+\sum_{i=k+1}^{N}\delta t_{k,i},
\ee
where
\be
\delta t_{k,i}=\left\{
\begin{aligned}
&2i\frac{\left|1-\xi_k\right|^2\left|1+\xi_k\right|^2}{(\xi_k^2-\bar\xi_k^2)}\log \left[\left|\frac{\xi_i-\xi_k}{\bar\xi_i-\xi_k}\right|^4 \left|\frac{\xi_i\xi_k-1}{\bar\xi_i\xi_k-1}\right|^2\right]&\text{if}\;\, {\bf \Omega}_k^\dagger {\bf \Omega}_i=1, \\
\\
&2i\frac{\left|1-\xi_k\right|^2\left|1+\xi_k\right|^2}{(\xi_k^2-\bar\xi_k^2)}\log \left[\left|\frac{\xi_i-\xi_k}{\bar\xi_i-\xi_k}\right|^2\right]&\text{if}\;\, {\bf \Omega}_k^\dagger {\bf \Omega}_i=0.
\end{aligned}\right.
\ee
These time delays correspond to type I and II scattering in the notations of \cite{Hatsuda:2009pc}, and the interpretation of the above result is that $N$-soliton scattering factorizes into a sequence of $2$-body interactions. Indeed, it can be shown that for the special cases where the polarizations are parallel or orthogonal, the interaction leaves their direction unaffected and hence we retrieve the familiar picture of soliton scattering \cite{1972JPSJ...33.1459H, Kalousios:2008gz}. As was explained in \cite{Manakov}, the fact that the story is more complicated for general polarizations, is precisely related to the change of their direction by the interaction.

\section{Conclusions and Open Questions}

In this paper we have constructed explicit scattering solutions of an arbitrary number of solitons $N$ for the sigma models with target spaces $CP^n$, $SU(n)$ and $S^n$. All the input that is required are the complex parameters $\xi_i$ and polarization vectors $e_i$, $i=1,\ldots,N$ characterizing each magnon, together with the auxiliary field corresponding to the vacuum $\Psi_0(\l)$, and then the solution can be calculated directly, without having to determine intermediate dressed quantities.

Given our interest in the study of integrability in the $AdS_5/CFT_4$ and $AdS_4/CFT_3$ correspondence, our choice of sigma models was based on the fact that they are equivalent to string theory on certain subspaces of the $AdS_5\times S^5$ and $AdS_4\times CP^3$ geometrical backgrounds respectively. As we explain in appendix \ref{Examples}, for particular choices of vacua we obtain giant magnon solutions, which are dual to excitations in the spin chain language of the corresponding CFT.

The success of our method for solving the recursion of the dressing method for these three cases, makes it natural to ask whether it could be generalized to other symmetric spaces as well. An important ingredient of our proof was that the quantity to be determined must be chosen so that its dressing factor contains a single, rank 1 projection operator $\cal P$, namely $\chi\sim 1+ c {\cal P}$, $Tr[{\cal P}]=1$, for some $c$. For example, in the $CP^n$ case, even though the matrix field $g$ has a complicated dressing factor, we were able to proceed by selecting the complex variable $Z$, whose dressing factor is indeed of the required form. Clearly the symmetric spaces with just one pole in the dressing matrix are the first candidates for satisfying the aforementioned rule, and there exist several of them \cite{Harnad:1983we}, at least in the minimal solution with $|\xi_i|=1$, such as $SU(n)/O(n)$ or $SU(2n)/Sp(n)$. For more poles, it is possible that instead of the matrix field $g$, another related quantity with simpler dressing matrix may be defined.

In this work, we have focused on Riemannian symmetric spaces with positive definite metric and compact size, for which the dressing method can be applied in a more transparent manner. Recently however, similar integrable methods for obtaining solutions have been successively used for noncompact cases such as $AdS$ space \cite{Jevicki:2007pk, Jevicki:2007aa, Jevicki:2008mm}. Given the usefulness of such methods in the programme of computing $\mathcal{N}=4$ SYM amplitudes at strong coupling \cite{Alday:2009yn} and the exciting new connections with integrability that it has led to \cite{Alday:2009dv, Alday:2010vh}, it would be worthwhile to study whether our results could be generalized for the $AdS$ space.

On a separate topic, in appendix \ref{Examples} we have derived the determinant formulas (\ref{Z_Nminus})-(\ref{Z_Nplus}) and (\ref{Z_N^1_in_SU(2)}) for certain components of the $CP^3$ and $SU(2)$ $N$-magnon solutions. Their form is very similar to the $N$-soliton solutions of the sine-Gordon theory which were used in \cite{Babelon:1993bx} in order to obtain an effective particle description encoded in the Ruijsenaars-Schneider model \cite{Ruijsenaars:1986vq}. It would thus be interesting to investigate whether our determinant formulas can be used analogously, as the necessity for deriving an effective Hamiltonian describing magnon dynamics was pointed out in \cite{Dorey:2007xn}. In particular, the transition from the soliton to the particle description is performed by expressing the Lax pair in terms of the eigenvalues of the matrices entering the determinant formulas, so a concrete task would be to check if this can be done in the magnon case. Perhaps the analogy can be made more precise by translating our solutions to the symmetric space sine-Gordon (SSSG) theory which arises from the Pohlmeyer reduction of the sigma model \cite{Pohlmeyer:1975nb}, according to the method of \cite{Hollowood:2009tw}.

\section*{Acknowledgments}
We would like to thank to A. Jevicki, K. Jin and A. Volovich for collaboration in related topics.  We are also grateful to L. Faddeev, A. V. Mikhailov and M. Spradlin for helpful discussions, suggestions as well as comments on the manuscript.  C.K. would like to thank ICTP for hospitality during the last stages of this work.  C.K. is supported by Deutsche Forschungsgemeinschaft via SFB 647.

\appendix

\section{Symmetric Spaces}\label{appx_symm_spaces}
A symmetric space is a quotient of two Lie groups $G/H$ with the following additional structure :
\begin{enumerate}
\item There exists a 1-1 map $\s_-$ from $G$ to itself that obeys
\be
\s_-(\s_-(g))=g,\quad \s_-(g g')=\s_-(g)\s_-(g')\quad \forall g,g'\in G.\label{involutive_automorphism}
\ee
Namely the map is its own inverse and preserves the structure of group multiplication, hence being described as an `involutive automorphism'.
\item The subgroup $H$ is fixed by demanding that all its elements are invariant under $\s_-$,
\be
\s_-(h)=h\quad \forall h\in H.
\ee
\end{enumerate}

There is a natural way to embed the quotient $G/H$ inside $G$, by mapping each of its elements, cosets $gH$, to the element $\tilde g\in G$
\be\label{cartan_immersion}
\tilde g=\s_-(g)g^{-1}.
\ee
Indeed, from the properties (\ref{involutive_automorphism}) it very easy to show that two elements in the same coset $g, g'$, in other words related with the equivalence relation $g'=gh$ for some $h\in H$, will be mapped to the same $\tilde g$ according to (\ref{cartan_immersion}).

Similarly, we may show that (\ref{cartan_immersion}) implies
\be\label{symmspace_constraint}
\sigma_-(\tilde g)=\tilde g^{-1},
\ee
and so we may think of $G/H$ as the submanifold $\Sigma$ spanned by all elements $\tilde g$ in $G$, subject to the constraint (\ref{symmspace_constraint}),
\be
\Sigma=\{\tilde g\in G|\; \s_-(\tilde g)=\tilde g^{-1}\}.
\ee

\section{Examples of Giant Magnons}\label{Examples}

In this appendix we work out in more detail the $N$-magnon solutions arising from the use of particular vacua and polarization vectors.

\subsection{$CP^3$}\label{appx_CP3}
As we mentioned in the main text, giant magnons in $CP^3$ are relevant for the more recent $AdS_4 /CFT_3$ correspondence. We choose as vacuum\footnote{Here we have specialized our gauge so that $\kappa=2$ in \eqref{Virasoro44}, in order to be in agreement with the conventions of \cite{Hollowood:2009sc}.}
\be
Z_0 = (\cos t~~ -\sin t ~~ 0 ~~ 0)^{\rm T},
\ee
which leads to the following solution to the linear system \eqref{linearsystem},
\be\label{cpPsi=}
\Psi_0(\l)=
\begin{pmatrix} \cos  \mathcal{Z}(\l) & -\sin  \mathcal{Z}(\l) & 0 & 0\\
\sin  \mathcal{Z}(\l) & \cos  \mathcal{Z}(\l) & 0 & 0 \\
0 & 0 & 1 & 0 \\
0 & 0 & 0 & 1
\end{pmatrix},
\ee
where
\be\label{calZ_in_CP}
\mathcal{Z}(\l)=2( \frac{x_+}{1+\l} +\frac{x_-}{1-\l}),
\ee
and $x_\pm=(t\pm x)/2$ as usual.
From \eqref{cpPsi=} it is easy to form $h_i$ and $\a_{ij}$ by direct substitution to \eqref{cph} and \eqref{cpalpha}, and then the $N$-magnon solution is given by (\ref{Znfinal}).

Let us illustrate how this procedure works by examining solutions consisting of the elementary dyonic magnons on $CP^2$ of \cite{Abbott:2009um, Hollowood:2009sc}. Namely we restrict the polarization vectors to be of the form $e_i=(1~~ i ~~ {\bf \Omega}_i)^{\rm T}$ where for the moment we consider the two-component vectors ${\bf \Omega}_i$ to be general. This implies that
\be
h_j=e^{-i {\cal \bar Z}_j}\left(\begin{array}{c}
-1\\
i\\
\vec 0
\end{array}\right)
+
\left(\begin{array}{c}
0\\
0\\
{\bf \Omega}_j
\end{array}\right), \quad h_i^\dagger Z_0=-e^{-it+i {\cal Z}_i }
\ee
and
\be
\a_{ij}=e^{i ({\cal Z}_i-{\cal \bar Z}_j)}\frac{-2\xi_i}{\xi_i-\bar\xi_j}+K_{ij},
\ee
where ${\cal Z}_i={\cal Z}(\xi_i), {\cal \bar Z}_i={\cal Z}(\bar\xi_i)$ and
\be
K_{ij}=\frac{(1-\xi_i^2)\bar \xi_j {\bf\Omega}^\dagger_i {\bf \Omega}_j}{(\xi_i-\bar \xi_j)(\xi_i\bar \xi_j-1)}.
\ee
Further simplifications occur if we take the following linear combinations of the first two vector components of $Z_N$, $Z_N^{\pm}=Z_N(1)\pm i Z_N(2)$, and similarly for all other vectors\footnote{
This is essentially the same with the basis of projective coordinates considered in \cite{Hatsuda:2009pc}. In particular, if we have a vector $v=(v_1~~ v_2~~ v_3~~ v_4)^T$ in our original basis, it is transformed into $v=(\frac{v_1-i v_2}{\sqrt{2}}~~v_3~~ v_4~~\frac{v_1+i v_2}{\sqrt{2}})^T$ in the basis of the latter reference.}
\be
Z_0^\pm=e^{\mp it},\quad h_j^+=-2e^{-i {\cal \bar Z}_j},\quad h_j^-=0,
\ee
in which case all elements of the first row of $Z_N^-$ in (\ref{Znfinal}) but $Z_0^-$ will be zero, and hence
\be\label{Z_Nminus}
Z_N^-=e^{it}\det (e^{i ({\cal Z}_i-{\cal \bar Z}_j)}\frac{-2\xi_i}{\xi_i-\bar\xi_j}+K_{ij}),\quad i,j=1,\ldots,N.
\ee
The $Z_N^+$ case is only slightly more complicated, as we may multiply the first row in (\ref{Znfinal}) with $e^{i {\cal Z}_i}$ and add it to the $i$-th row for $i=2,\ldots,N$, which now makes all elements of the first column but $Z_0^+$ zero, and hence yields
\be\label{Z_Nplus}
Z_N^+=e^{-it}\det (e^{i ({\cal Z}_i-{\cal \bar Z}_j)}\frac{-2\bar\xi_j}{\xi_i-\bar\xi_j}+K_{ij}),\quad i,j=1,\ldots,N.
\ee
Finally, the remaining two components of $Z_N$, which we shall denote as $Z_N^\perp$, will be given by
\be\label{Zperp}
Z_N^\perp=e^{-it}\left|
\begin{array}{c|ccc}
0&&{\bf \Omega}_j&\\\hline
&&&\\
e^{i {\cal Z}_i}&&-2e^{i ({\cal Z}_i-{\cal \bar Z}_j)}\frac{\xi_i}{\xi_i-\bar\xi_j}+K_{ij}&\\
&&&\\
\end{array}\right|,
\ee
where we have denoted the elements of the $(1\times N), (N\times 1)$ and $(N\times N)$ submatrices. In our conventions, the single elementary dyonic magnon will thus be given by
\be\label{Z_1}
\begin{aligned}
Z_N^-&=e^{it}(e^{i ({\cal Z}_1-{\cal \bar Z}_1)}\frac{-2\xi_1}{\xi_1-\bar\xi_1}+K_{11}),\\
Z_N^+&=e^{-it} (e^{i ({\cal Z}_1-{\cal \bar Z}_1)}\frac{-2\bar\xi_1}{\xi_1-\bar\xi_1}+K_{11}),\\
Z_N^\perp&=-e^{-it}e^{i {\cal Z}_1}{\bf \Omega}_1.
\end{aligned}
\ee

\subsection{$SU(2)$}\label{appSU2}
Let us start by reviewing the notations and conventions of \cite{Spradlin:2006wk}, where the dressing method was first used in the context of giant magnons, reproducing the dyonic solution of \cite{Chen:2006gea} and providing a variety of new solutions.

String theory on $R \times S^3$ in conformal gauge admits the same equations of motion as the $SU(2)$ principal chiral model through the embedding
\be\label{SU(2)_embedding}
g=\begin{pmatrix}  Z^1 & -i Z^2  \\ -i \bar{Z^2} & \bar{Z^1} \end{pmatrix},
\ee
where $Z^1 =X^1 + i X^2, ~Z^2 =X^3+i X^4$ and $X^i$ the spacetime coordinates of the string theory on $R \times S^3$ such that $\sum_i (X^i)^2 =1$.

We can choose as a vacuum a point-like string that moves around the equator of $S^3$ with the speed of light, namely $Z^1 = e^{it},~Z^2=0$.  Then the corresponding vacuum matrices $g_0$ and $\Psi_0(\l)$ are given by
\be
g_0=\begin{pmatrix}  e^{it} & 0  \\ 0 & e^{-it} \end{pmatrix}, \qquad
\Psi_0(\l)=\begin{pmatrix}  e^{i \mathcal{Z}(\l)} & 0  \\ 0 &    e^{-i \mathcal{Z}(\l)} \end{pmatrix}
\ee
where now $\mathcal{Z}(\l)$ is
\be\label{Z=}
\mathcal{Z}(\l)= \frac{x_+}{1+\l} +\frac{x_-}{1-\l}.
\ee
One can easily check that the matrix $\Psi_0(\l)$ satisfies the constraint \eqref{constraint44}.  As we can see from (\ref{SU(2)_embedding}), the first and second column of $g$ in fact contains the same amount of information, all we need to keep track of is the quantity
\be
Z=\begin{pmatrix}g_{11} \\ g_{21}\end{pmatrix}=\begin{pmatrix}  Z^1 \\ -i \bar{Z^2}\end{pmatrix}
\ee
instead. Then, in line with footnote \ref{footnote_SU(n)}, the numerator of the $N$-magnon solution (\ref{gnDet}) reduces to (\ref{Znfinal}), where now
\be
\begin{aligned}
Z_0&=\begin{pmatrix}  e^{it}  \\ 0  \end{pmatrix}&h_j&=\begin{pmatrix} e^{+i \mathcal{\bar Z}_j} e_j^1 \\ e^{-i \mathcal{Z}_j } e_j^2 \end{pmatrix}\\
h_i^\dagger Z_0&=e^{it-i {\cal Z}_i }\bar e^1_i\quad&\a_{ij}&=-\frac{\xi_i}{\xi_i-\bar{\xi}_j} \left( e^{-i (\mathcal{Z}_i- \mathcal{\bar Z}_j) } \bar{e}_i^1 e_j^1 +e^{+i (\mathcal{Z}_i- \mathcal{\bar Z}_j) } \bar{e}_i^2 e_j^2   \right),
\end{aligned}
\ee
with the $i$-th magnon having polarization vector $e_i = (e_i^1 ~ e_i^2)^{\rm T}$. Particularly for the first component of the $N$-soliton column vector $Z_N$, determinant manipulations similar to the ones used to obtain (\ref{Z_Nplus}), yield
\be\label{Z_N^1_in_SU(2)}
Z_N^1=\prod_{i=1}^N \left(\frac{\xi_i}{\bar\xi_i}\right)^{1/2} e^{it}\;\frac{\det(\a_{ij}+e^{-i (\mathcal{Z}_i- \mathcal{\bar Z}_j) } \bar{e}_i^1 e_j^1)}{\det(\a_{ij})}\;,\quad i,j=1,\ldots, N.
\ee
The product of $\sqrt{\xi_i/\bar\xi_i}$ comes from the transition from $U(n)$ to $SU(n)$  that we describe in the end of section \ref{subsection_SU(n)}.

The particular case where all $e_i=(1 ~ 1)^{\rm T}$ was discussed in \cite{Kalousios:2008gz}, and it is straightforward to show that for the given choice of polarizations, our current, more general results agree with the ones presented in the latter paper.

\subsection{$S^2$}
To apply the dressing method for strings moving in $R\times S^2$ we start by considering as vacuum a point particle that moves on the equator of the 2-sphere according to
\be
X_0 = (\cos t~~ \sin t ~~ 0)^{\rm T}.
\ee
Then the solution to the linear system \eqref{linearsystem} gives us
\be\label{SnPsi}
\Psi_0(\l)=
\begin{pmatrix} \cos 2 \mathcal{Z}(\l) & \sin 2 \mathcal{Z}(\l) & 0 \\
-\sin 2 \mathcal{Z}(\l) & \cos 2 \mathcal{Z}(\l) & 0 \\
0 & 0 & 1
\end{pmatrix},
\ee
where $\mathcal{Z}(\l)$ is given as previously from \eqref{Z=}.  Knowledge of $\Psi_0(\l)$ can easily give us $h_i$ and $\a_{ij}$ by direct substitution to \eqref{sh=} and \eqref{salpha=}, and the $N$-magnon solution is given by \eqref{Xn=}. Clearly, the relation between the $CP^n$ and $S^n$ model, which we described in section \ref{subsection_SU(n)}, together with the essentially identical choices of vacua (\ref{cpPsi=}) and (\ref{SnPsi}), up to different conventions, allow us to obtain $S^n$ giant magnons from slight modifications to the $CP^n$ ones.

\end{document}